# A Phase-Field-Micromechanics Study on the Microstructural Evolution during Viscous Sintering

Xiaoxu Dai[1], Bo Qian[1], Arkadz Kirshtein[2], Qingcheng Yang[1,3]*

[1] Shanghai Key Laboratory of Mechanics in Energy Engineering, School of Mechanics and Engineering Science, Shanghai Institute of Applied Mathematics and Mechanics, Shanghai University, Shanghai, China

[2] Department of Mathematics & Statistics, Texas A&M University-Corpus Christi, Corpus Christi, TX 78412, USA

[3] Shanghai Institute of Aircraft Mechanics and Control, Shanghai, China

* Corresponding author, Email: qyang@shu.edu.cn, Tel: 86-18116136627.

# Abstract

In the manufacturing process of high-performance particulate materials, viscous sintering plays a crucial role, particularly in fields such as polymer processing and additive manufacturing. The interactions between microscopic particles, their flow behavior, and the evolution of porosity during the viscous sintering process directly influence the material's density and mechanical properties. Therefore, developing efficient modeling techniques to simulate the viscous sintering process is essential for optimizing sintering technology. However, the large deformations and dynamic surface evolution inherent in the viscous sintering of particulate materials present challenges to traditional methods based on the sharp interface model. To address these challenges, we propose a thermodynamically consistent diffusion interface model, referred to as the phase-field-micromechanics model, to analyze the evolution of various physical quantities throughout the viscous sintering process. This model implicitly describes the evolution of particle morphology through an introduced phase-field variable. Through comparisons with analytical solutions and experimental data, we rigorously validate the correctness of the proposed model qualitatively and quantitatively under both isothermal and non-isothermal conditions. Using the proposed model, we explore the development of strain and stress during the sintering process, as well as the effects of particle size, shape and arrangement on the overall sintering behavior. The evolution of these characteristic indicators allows for a clear observation of the viscous sintering process, which is vital for understanding the mechanisms behind viscous sintering and for guiding industrial production.

# 1. Introduction

Sintering is a process driven by surface tension that involves expanding the contact area between powder particles and filling the pores or gaps between them under appropriate temperature, pressure, and environmental conditions[1]. Sintering is primarily classified into three types: solid-state sintering of crystalline materials, viscous sintering of amorphous materials, and liquid-phase sintering of crystalline materials[2]. The viscous sintering of amorphous particles is a phenomenon commonly observed in nature[3] as well as in industrial production [4–7], and it has been applied in the manufacture of various materials, including ceramics, plastics, glass, and polymer powders.

Today, viscous sintering is not only an indispensable component of many material processing workflows but also plays an increasingly significant role in the production of new building materials[8,9], novel batteries[10], pharmaceuticals[11], biomedical applications[12], high-precision processing[7], and aerospace[13]. Particularly in the field of additive manufacturing (3D printing) [2,6,14–17], a variety of materials have been utilized in the 3D printing process. Technological advancements have led to a growing demand for new materials, underscoring the need for a deeper understanding of sintering mechanisms and the accompanying microstructural evolution. Therefore, the research and application of sintering technology hold profound significance and importance for future process improvements and the production of innovative high-performance materials.

Due to geometric constraints and limitations in multi-particle modeling, most existing studies on the viscous sintering process focus on the sintering of cylindrical or sphere particles. In the realm of analytical models for two circular particles, Frenkel was the first to propose an analytical model for the shape evolution of two spherical particles during coalescence[18]. He conceptualized viscous sintering as a two-stage process. In the initial stage, the surface area of the contact region between neighboring particles expands until the interstitial pores are completely isolated. During the subsequent stage, these pores are gradually filled. Frenkel's approximation of neck growth was later refined by Eshelby, while Jacota, utilizing finite element analysis, demonstrated that Frenkel's model overestimated the neck growth rate [19]. Pokluda et al. advanced the corrected model by factoring in variations in particle radius [20]. Hopper, on the other hand, presented an analytical solution capturing the time-dependent evolution of viscous planar flow driven by surface tension within regions delineated by smooth, closed curves [21].

Numerical simulations of viscous sintering have utilized various methods, including the finite element method (FEM) [22–27], Monte Carlo methods[28,29], discrete element methods (DEM) [7,30,31], boundary element methods (BEM) [32,33], the Lattice Boltzmann Method (LBM) [34,35], and phase field methods (PFM) [36]. Wakai proposed a model for the sintering force of two spherical particles and found that the sintering force can be evaluated from the neck growth rate, and extended this model to simulations involving different shapes and three particles[25,26]. Polychronopoulos et al. developed a coalescence model for viscous particles of varying sizes, proposing a general and unified form of a surface tension-driven coalescence model. They further created a model to analyze pore changes during the sintering process, comparing it with analytical solutions and applying it to the manufacturing of PLA filaments via fused filament fabrication (FFF) [24,37,38]. Bordère et al. employed the Monte Carlo method to simulate the sintering of glass cylinders, comparing their findings with experimental data and discovering that the particles exhibited characteristics of non-Newtonian viscous flow in the later stages of sintering[29]. Balemans et al. introduced a sharp-interface model, which they validated by comparing its results with the analytical solution for the sintering of two equally sized particles. This model was subsequently applied to the SLS laser 3D printing process, where they investigated key parameters influencing sintering outcomes [22,39]. Leveraging a DEM-CFD-FEM coupled model, Xiao et al. delved into the mechanisms of particle viscous flow, interface evolution, and thermal stress distribution during the sintering process. Their study highlighted the superior mechanical properties of octahedral diamonds, attributed to their stable geometric structure, thereby laying a theoretical foundation for the efficient fabrication of high-performance diamond grinding wheels[7].

The sharp interface model has been widely utilized in viscous sintering simulations[22,24–26,37,38]. However, several limitations arise during the simulation processes inherent in the sharp interface approach. In the numerical implementation of these models, accurately and explicitly describing complex three-dimensional (3D) powder densification microstructures becomes increasingly challenging. The primary distinction between the phase field model and the sharp interface approach is its compelling feature: the phase field method introduces phase field variables that can naturally and implicitly capture the structural evolution during viscous sintering, thereby eliminating the need for explicit tracking of the continuously evolving microstructural features. As a result, the phase field method has emerged as one of the most powerful computational techniques for simulating the microstructural evolution across a variety of material processes.

Currently, the phase field method has been widely applied to simulate microstructural evolution in both solid-state sintering [40–60] and liquid-phase sintering[61–64], while its application in the context of viscous sintering remains relatively limited. Harnessing the advantages of the phase field methodology, we extend our previous phase field model specifically for viscous sintering from microstructures with uniform viscosity to systems with high viscosity ratios under both isothermal and non-isothermal conditions. The extended model is grounded in the principles of energy variational formulation. Consequently, the governing equations utilized can be derived consistently in accordance with energy conservation laws and the principle of least action. The extended model is validated qualitatively and quantitatively by comparing it with analytical and experimental results under both isothermal and non-isothermal conditions. The analysis investigates the development of strain and stress during the sintering process, as well as the effects of particle size, shape, and arrangement on overall sintering behavior.

The remainder of this manuscript is structured as follows: Section 2 introduces the proposed phase-field-micromechanics model for viscous sintering. In Section 3 we validate the proposed model both qualitatively and quantitatively through comparisons with theoretical analyses and experimental results under isothermal and non-isothermal conditions. Section 4 presents a straightforward application of the proposed model to demonstrate its capability in capturing the evolution of particle morphology, stress and strain during the sintering process, as well as the influence of particle size and rearrangement on viscous sintering behavior. Finally, Section 5 provides concluding remarks on the model's performance and outlines potential directions for future research.

## 2. Theory

2.1. Governing equations

In the phase-field method, microstructural evolution is captured through a set of phase-field variables, which are continuous functions of time and spatial coordinates. Phase-field variables are typically classified into two categories: conserved variables and non-conserved variables (or order parameters). Conserved order parameters are usually related to local composition and must satisfy local conservation conditions. The evolution of conserved variables is described by the Cahn-Hilliard diffusion equation. Non-conserved variables generally contain information about local material structure and orientation. The evolution of non-conserved order parameters is governed by the Allen-Cahn equation.

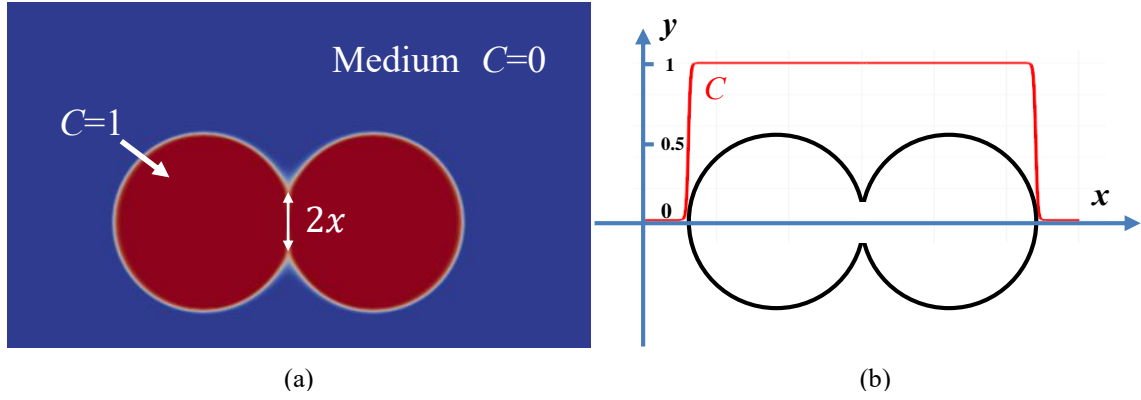

(a)                  (b)

**Fig. 1.** Definition of the introduced phase-field variable $C$: (a) An schematic representation of the phase-field variable $C$; (b) Spatial distribution of the phase-field variable $C$ (red line) along the $x$-axis connecting the particle centers.

For viscous sintering, we introduce a phase-field variable $C$ to distinguish the particle phase from the surrounding vapor medium (as shown in Fig. 1a), where $C=1$ within the particles and $C=0$ in the surrounding medium, with a rapid change occurring at the particle-vapor interface (as illustrated in Fig. 1b). Thus, the evolution of $C$ allows us to implicitly observe the evolution of the particle morphology. Assuming a constant sintering temperature, the total free energy $\mathcal{F}$ of the system can be expressed in terms of surface energy: $\mathcal{F} = \mathcal{F}_{sf}$, which can be defined as a function of the introduced phase-field $C$ as follows:

$$\mathcal{F}_{sf} = \int_{\Omega} \left( \alpha C^2 (1-C)^2 + \frac{\kappa_C}{2} \|\nabla C\|^2 \right) d\Omega \quad (1)$$

where $\alpha$ and $\kappa_C$ are model parameters related to the actual material properties. Let $\gamma^{sf}$

represent the surface tension of the material and $\delta^{sf}$ the surface thickness of the particles, $\alpha$ and $\kappa_C$ can be expressed as follows:

$$\begin{cases} \gamma^{sf} = \frac{\sqrt{2\kappa_C \alpha}}{6} \\ \delta^{sf} = \sqrt{\frac{8\kappa_C}{\alpha}} \end{cases} \text{ or } \begin{cases} \alpha = 12\frac{\gamma^{sf}}{\delta^{sf}} \\ \kappa_C = \frac{3}{2}\gamma^{sf}\delta^{sf} \end{cases} \quad (2)$$

The detailed derivation of Eq. (2) can be found in [56]. Assuming the particles are incompressible, the governing equations for viscous sintering can be derived using the energy variational method and are given as follows [65]:

$$\frac{\partial C}{\partial t} + \nabla \cdot (C\dot{\mathbf{u}}) = \nabla \cdot \left(M\nabla \frac{\delta \mathcal{F}}{\delta C}\right) \quad (3)$$

$$\nabla \cdot \dot{\mathbf{u}} = 0 \quad (4)$$

$$\nabla \cdot \boldsymbol{\sigma} + \mathbf{b} = 0 \quad (5)$$

where Eq. (3) is the modified Cahn-Hilliard equation that includes a convection term, $M$ is a small relaxation parameter, and $\dot{\mathbf{u}}$ represents the mechanics velocity; Eq. (4) represents the incompressibility condition and the mechanical equilibrium of the entire system is denoted by Eq. (5), where the stress tensor $\boldsymbol{\sigma}$ and the body force term $\mathbf{b}$ are given as:

$$\boldsymbol{\sigma} = \mu^{eff}(\nabla \dot{\mathbf{u}} + (\nabla \dot{\mathbf{u}})^T) + p_L \mathbf{I} \quad (6)$$

$$\mathbf{b} = -\left(\nabla \cdot \left(\frac{\partial f}{\partial \nabla C} \otimes \nabla C\right) - \nabla f\right) \quad (7)$$

where $f$ is the free energy density and is defined as: $f = \alpha C^2 (1-C)^2 + \frac{\kappa_C}{2}\|\nabla C\|^2$. In Eq. (6), $p_L$ denotes pressure and $\mu^{eff}$ is the effective viscosity. Let $\eta_p$ and $\eta_v$ represent the viscosities of the material and the surrounding medium respectively, and $\beta$ be the ratio of $\eta_v$ to $\eta_p$. The effective viscosity $\eta^{eff}$ of the system is defined as:

$$\eta^{eff} = \eta_p N(C) + \eta_v[1 - N(C)] = (\beta + (1-\beta)N(C))\eta_p \quad (8)$$

where $N(C)$ is a constructed interpolation function of $C$, and is defined as $N(C) = C^2[1 + 2(1-C) + \omega(1-C)^2], \omega > 3$. In this work, $\beta$ takes a value of 0.001. Furthermore, let $p = p_L + f$, then Eq. (5) can be rewritten as:

$$\nabla \cdot [\eta^{eff}(\nabla \dot{\mathbf{u}} + (\nabla \dot{\mathbf{u}})^T)] + \nabla p - \nabla \cdot \left(\frac{\partial f}{\partial \nabla C} \otimes \nabla C\right) = 0 \quad (9)$$

2.2. Weak forms of the governing equations

In this section, the weak forms of the correspond governing Eqs. (3),(4), and (9) are derived in order to use the Finite Element Method as the partial differential equation solver.

To employ the $C_0$-finite element, the fourth-order Eq. (3) is splitted as two second-order partial differential equations:

$$\frac{\partial C}{\partial t} + \nabla \cdot (C\dot{\mathbf{u}}) = \nabla \cdot (M\nabla \mu_C) \tag{10}$$

$$\mu_C = \frac{\delta \mathcal{F}}{\delta C} = g(C) - \kappa_C \nabla^2 C \tag{11}$$

where $g(C)$ is defined as: $g(C) = 2\alpha C(1-C)(1-2C)$. Applying integration by parts and Gauss's theorem, we can obtain the weak forms of the correspond governing equations as follows:

$$\int_\Omega \frac{\partial C}{\partial t} v_C d\Omega + \int_\Omega M\nabla \mu \cdot \nabla v_C \, d\Omega - \int_\Omega (C\dot{\mathbf{u}}) \cdot \nabla v_C \, d\Omega = 0 \tag{12}$$

$$\int_\Omega \mu_C v_\mu d\Omega - \int_\Omega \kappa_C \nabla C \cdot \nabla v_\mu \, d\Omega - \int_\Omega g(C) v_\mu \, d\Omega = 0 \tag{13}$$

$$\int_\Omega \left( \eta^{eff} (\nabla \dot{\mathbf{u}} + (\nabla \dot{\mathbf{u}})^T) \right) : \nabla \mathbf{v}_{\dot{\mathbf{u}}} d\Omega + \int_\Omega p(\nabla \cdot \mathbf{v}_{\dot{\mathbf{u}}}) d\Omega -$$
$$\int_\Omega (\kappa_C \nabla C \otimes \nabla C) : \nabla \mathbf{v}_{\dot{\mathbf{u}}} d\Omega = 0 \tag{14}$$

$$\int_\Omega (\nabla \cdot \dot{\mathbf{u}}) v_p d\Omega = 0 \tag{15}$$

where $v_C, v_\mu, \mathbf{v}_{\dot{\mathbf{u}}}$, and $v_p$ are the test functions for the unknown variables $C$, $\mu$, $\dot{\mathbf{u}}$, and $p$, respectively. In deriving the weak forms, the employed boundary conditions are as follows:

$$M\nabla \mu_C \cdot \mathbf{n} = 0 \tag{16}$$

$$\kappa_C \nabla C \cdot \mathbf{n} = 0 \tag{17}$$

$$\dot{\mathbf{u}} = 0 \tag{18}$$

where $\mathbf{n}$ represents the outward normal vector at the boundary.

## 3. Model Validation

### 3.1. The employed materials and computational settings

To validate the proposed phase-field-micromechanics model in Section 2, we selected the viscous sintering of two identical circular particles in two dimensions (2D) as a benchmark. The correctness of the model is verified by comparing the growth of the neck formed at the contact point between the two particles (as illustrated in Fig. 1). We chose two materials: PA12 and ABS. Table 1 and Table 2 present the originally employed material parameters for PA12 and ABS, respectively. We utilized the open-source FEniCS platform to solve the governing equations in Section 2 using the finite element method.

**Table 1.** Material properties of polyamide 12 (PA12) powder at $T = 175\,°C$ [66].

| Parameter | Symbol | Value |
|---|---|---|
| Viscosity(Pa·s) | $\mu_p$ | 400 |
| Surface tension (N/m) | $\gamma^{sf}$ | 0.03 |
| Initial particle radius(μm) | $R_0$ | 30 |

**Table 2.** Material properties of ABS powder at $T = 240\,°C$ [67].

| Parameter | Symbol | Value |
|---|---|---|
| Viscosity(Pa·s) | $\mu_p$ | 5100 |
| Surface tension (N/m) | $\gamma^{sf}$ | 0.029 |
| Initial particle radius(μm) | $R_0$ | 235 |

Using PA12 material ($T = 175°C$) as an example, the surface thickness of the two circular particles was set to $\delta^{sf} = 3h$, where $h$ represents the distance between two spatial mesh points. Unless specified otherwise, this surface thickness will be maintained in subsequent examples. To eliminate the influence of mesh accuracy on the model's performance, we selected three different mesh densities represented as M1, M2 and M3, respectively, and the dimension of each grid is depicted in Table 3. Let $x$ denote the contact radius (as shown in Fig. 1) and $R_0$ be the initial particle radius. Throughout the sintering process of the two circles, the total area remains constant, and the radius of the larger circle formed at the conclusion of sintering is $\sqrt{2}R_0$, as depicted in Fig. 2, where the horizontal axis is the normalized time, defined as $t^* = \gamma^{sf} t/R_0\mu_p$, while the vertical axis represents the normalized contact radius, $X^* = x/\sqrt{2}R_0$. The three curves from the employed meshes in Fig. 2 nearly overlap, indicating that the mesh size of M1 achieves the desired accuracy.

**Table 3.** Mesh resolution along each dimension employed in the convergence study.

| Mesh | Number of Elements |
|---|---|
| M1 | 240×180 |
| M2 | 480×360 |
| M3 | 960×720 |

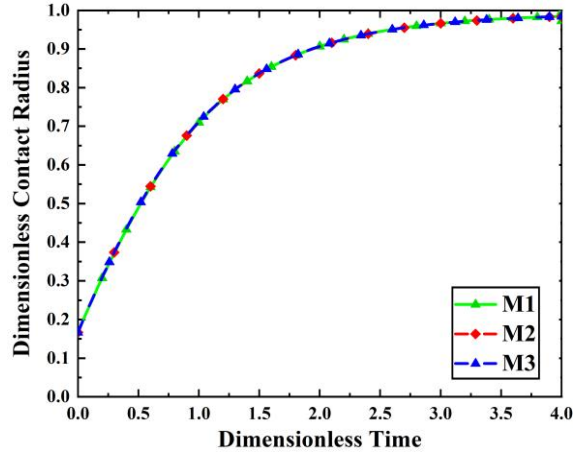

**Fig. 2.** The evolution of the normalized neck length as a function of normalized time during the sintering of the employed two-particle model with three different mesh resolutions.

3.2. Comparison to analytical results

Employing the material PA12, Table 4 presents the normalized model parameters according to the map between material parameters and model parameters defined in Eq. (2). Under the assumption of constant temperature, the normalized contact radius was compared with Hopper's analytical results[68]. Hopper's solution describes the temporal evolution of creeping viscous, incompressible planar flow within a finite domain, with one specific shape offering an exact solution for the coalescence of two identical cylinders[21], which provides a robust foundation for validating the model's accuracy.

**Table 4.** The normalized model parameters for PA12 and ABS materials.

| Materials | $\alpha^*$ | $\kappa_C^*$ | $\mu_p^*$ | $R_0^*$ |
|---|---|---|---|---|
| PA12 ($T = 175°C$) | 120 | 135 | 0.4 | 30 |
| ABS ($T = 240°C$) | 120 | 60 | 34.94 | 23.5 |

Fig. 3 illustrates the two-dimensional microstructural evolution of two equally sized circular particles. As the sintering process advances, the particles progressively draw closer and fused to form a larger circular structure. Fig. 4 illustrates the changes in the contact radius throughout the sintering process. The black line represents the analytical results, while the red line is from the proposed phase-field-micromechanics model. In general, the agreement is satisfactory, effectively validating the accuracy of the proposed model with sharp viscosity change across the particle-vapor interface. The observed slight difference in Fig. 4 may be induced by the diffuse-interface approximation of Hopper's sharp-interface solution.

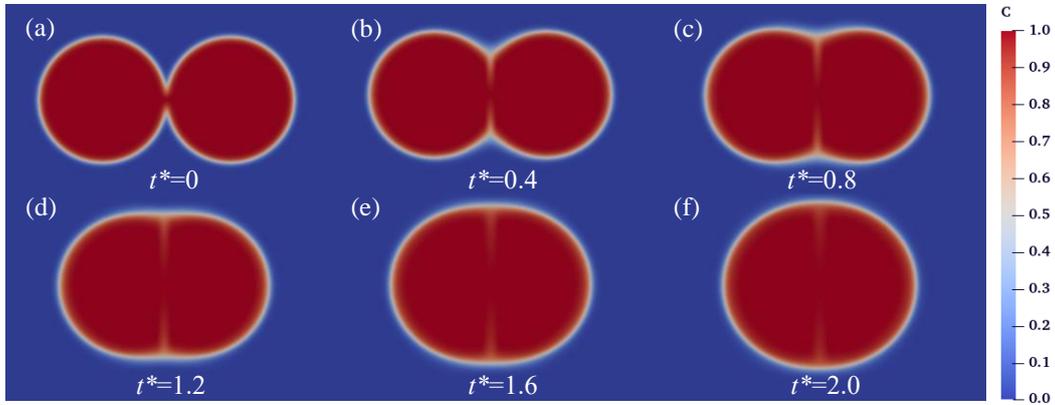

**Fig. 3.** Microstructure evolution of the two equally-sized particles with radius $R_0^*$ at (a) $t^*=0$, (b) $t^*=0.4$, (c) $t^*=0.8$, (d) $t^*=1.2$, (e) $t^*=1.6$, and (f) $t^*=2.0$. The $C$ in the color bar represents the introduced phase-field variable $C$ having a value of 1 within particle compact and of 0 within the surrounding medium represented by blue color (See the color bar).

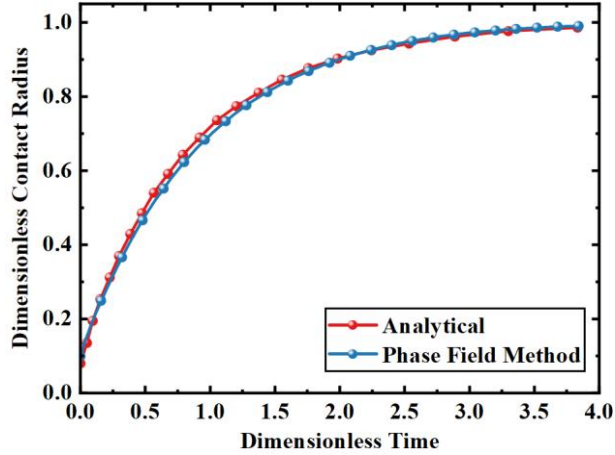

**Fig. 4.** The contact radius evolution from the respective Hopper's solution (the red curve) and the proposed phase-field-micromechanics model (the blue curve) for the sintering of the employed two PA12 particles.

3.3. Qualitative and quantitative experimental validation

Using ABS material at $T$=200°C, we first conducted a qualitative comparison with the experimental observations from [69]. Note that particle shapes are idealized in the propose phase-field-micromechanics model. Fig. 5a shows the experimentally observed microstructure evolution of two ABS particles at different sintering time under 200 °C, and Fig. 5b presents the particle morphology evolution predicted from the proposed model at the corresponding sintering times, where a general alignment with experiment observations is reached, indicating that the proposed model effectively captures the coalescence phenomenon in the polymer sintering process.

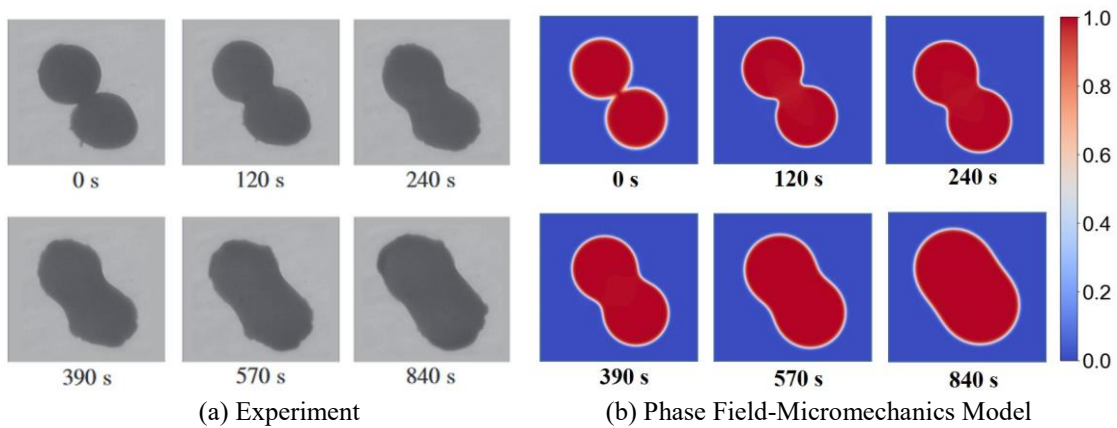

(a) Experiment     (b) Phase Field-Micromechanics Model

**Fig. 5.** Microstructure evolution during viscous sintering of two extruded ABS filaments at $T$=200℃ from experimental observations (a) and the proposed phase-field-micromechanics model (b).

To quantitatively validate the proposed model, we then compared the evolution of contact radius during the sintering of two ABS particles under isothermal condition at $T$=240°C from [69]. As shown in Fig. 6, the contact radius growth from the proposed model matches well with that from experimental data, indicating the capability of the proposed model to capture the sintering characteristics of real materials.

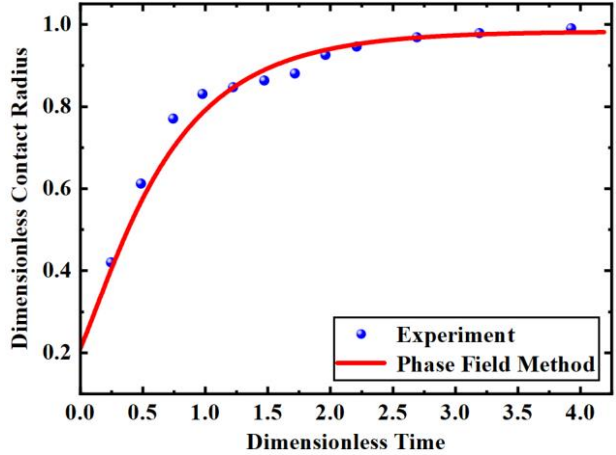

**Fig. 6.** The dimensionless contact radius growth profile of two ABS particles during isothermal sintering at 240°C from experimental measurements (blue dots) and the proposed phase-field-micromechanics model (the red curve).

3.4. Quantitative experimental validation under non-isothermal conditions.

The non-isothermal simulations were conducted for ABS particles under two different heating rates, specifically 10°C and 15°C per minute, starting from 190°C. The temperature increases linearly with each time step, resulting in a uniform heating curve. The temperature-dependent parameters for the ABS material are provided in [67] and are presented in Table 5. The initial dimensionless neck radius was set to 0.3 in the simulation in order to align with the initially observed experimental data [67].

**Table 5.** Temperature-dependent material parameter for ABS materials [69].

| Parameter | Symbol | Value |
|---|---|---|
| Viscosity(Pa·s) | $\mu_p$ | $5100\exp^{(-0.056(T-513))}$ |
| Surface tension (N/m) | $\gamma^{sf}$ | $0.029-0.000345(T-513)$ |

Fig. 7 compares the predictions from the proposed phase-field-micromechanics model with the corresponding experimental measurements, where a reasonable agreement is reached in general. On the one hand, the reduction in surface tension at higher temperature reduces the driving force for sintering; on the other hand, the decrease in viscosity at higher temperature reduces the resistance to viscous flow and thus accelerates the viscous-flow-assisted mass transport. Since the deduction in viscosity is exponential and the reduction in surface tension in linear as temperature increases (see Table 5), the ABS particles with a heating rate of 15°C per minute exhibits faster sintering behavior than the one with a heating rate of 10°C per minute, which is well captured in the proposed model.

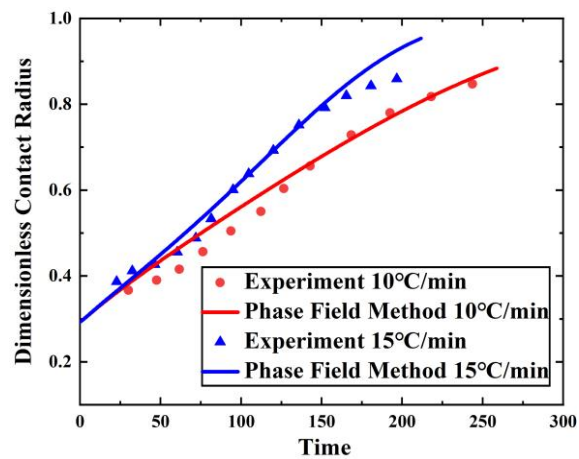

**Fig. 7.** The evolutions of dimensionless contact radius for ABS particles under heating rates of 10°C/min (red) and 15°C/min (blue), respectively, from experimental measurements (blue and red dots) and the proposed phase-field-micromechanics model (blue and red curves).

## 4. Numerical Examples

The validity of the proposed phase-field-micromechanics model is examined by its qualitative and quantitative performance under both isothermal and non-isothermal scenarios. In this section, we will delve into the factors shaping the viscous sintering process through numerical simulations and offer insights into the observed phenomena.

4.1. The influence of aspect ratio

In previous simulations, particles of uniform size were used. However, in real particle systems, coalescing particles often differ in size. To examine how varying radii influence the sintering process, particles of different radii under viscous sintering are investigated and analyzed. Fig. 8 provides a schematic of two unequally sized circles. At $t^*=0$, the two particles make contact at the origin of the coordinate system. Here, the initial radius of the smaller particle is $R_1$ and that of the larger particle is $R_2$. The initial size ratio of the particles is defined as: $b=R_2/R_1\geq 1$.

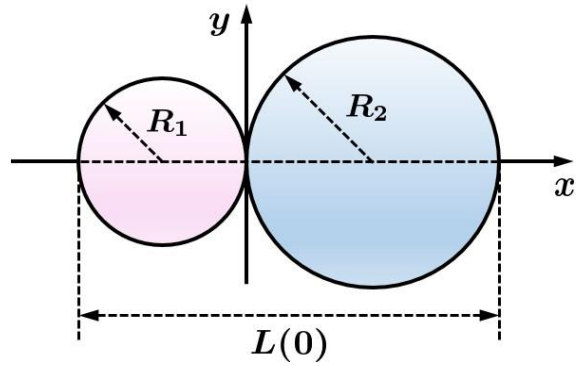

**Fig. 8.** Schematic diagram of the two unequally sized circles

As time progresses, a neck with a radius of $x$ forms at the intersection of the two particles. During the sintering process, these particles gradually coalesce, and we characterize the influence of size ratio $b$ on sintering through the evolution of the contact radius and the strain along the $x$-axis. Taking PA12 ($T=240°C$) as an example, Fig. 9a illustrates the growth of the neck radius across different $b$ values. Note that the normalized contact radius and time are consistently referenced to the smaller radius $R_1$.

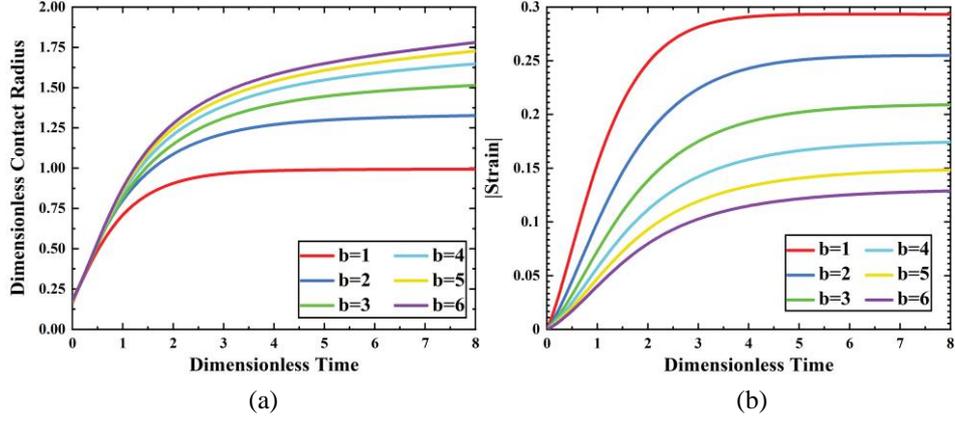

(a)                          (b)

**Fig. 9.** (a)The evolution of normalized contact radius as a function of normalized time under different size ratios *b*; (b)The *x*-strain evolution during the sintering of two particles with different size ratios.

In the early stages of sintering ($t^* \leq 0.5$), the impact of unequal size ratios on neck radius growth is minimal. However, beyond this point, the varying size ratios significantly influence the neck radius growth. For the case where *b*=1, complete sintering of the particles occurs around $t^*$=2.5. As *b* increases, the time required for complete sintering is delayed. This observation aligns with the findings of Polychronopoulos et al.[37], where the overall sintering time increases with the rise of the size ratio *b*.

Interestingly, while larger particles exhibit a greater contact radius, the differences among the curves diminish with increasing *b*. With the growth in size ratio, its effect on the sintering rate becomes progressively less significant, as the sintering process becomes increasingly dominated by larger particles, reducing the impact of smaller particles.

Fig. 9b illustrates the strain evolution during the sintering process for different size ratios. The absolute value of the shrinkage strain along the *x*-axis is defined as: $|\epsilon(t)| = \frac{|L(t^*) - L(0)|}{L(0)}$, where *L(0)* represents the initial length (as shown in Fig. 8) and *L(t\*)* denotes its length at time *t\**. As *b* increases, the rate of change of the strain curve slows, and the curve flattens more gradually. This further suggests that the influence of the increasing size ratio on sintering becomes less significant, and strain will increasingly depend on changes in the larger particles.

Additionally, it is observed that for the same *b* value, the rate of change along the *x*-direction slows notably after $t^*=3$, whereas the contact radius undergoes a rapid decline in its rate of change after $t^*=2$. This highlights that the neck region develops at a faster pace, with the *x*-direction continuing to contract for some time even as the neck growth nears completion.

4.2. Sintering of particle chains

Particles chains consisting of 2, 4, 6, 8, and 10 particles are considered respectively, with the evolution of the 4-particle chain over time is depicted in Fig. 10. Unlike solid-state sintering, which may stall at certain points, viscous sintering is free of grain boundaries, enabling multi-particle viscous sintering to progress to its final stage, ultimately resulting in the formation of a larger particle. The strain evolutions of each particle chain model are illustrated in Fig. 11. For a given curve, the strain rate displays a decreasing trend over time. Due to the fewer number of particles, the strain evolution observed during the sintering of two particles is more distinct compared to longer particle chains. This is attributed to the diminishing influence of the chain ends with an increasing number of particles, leading to a uniform mechanical environment for interior particles, and resulting in strain convergence.

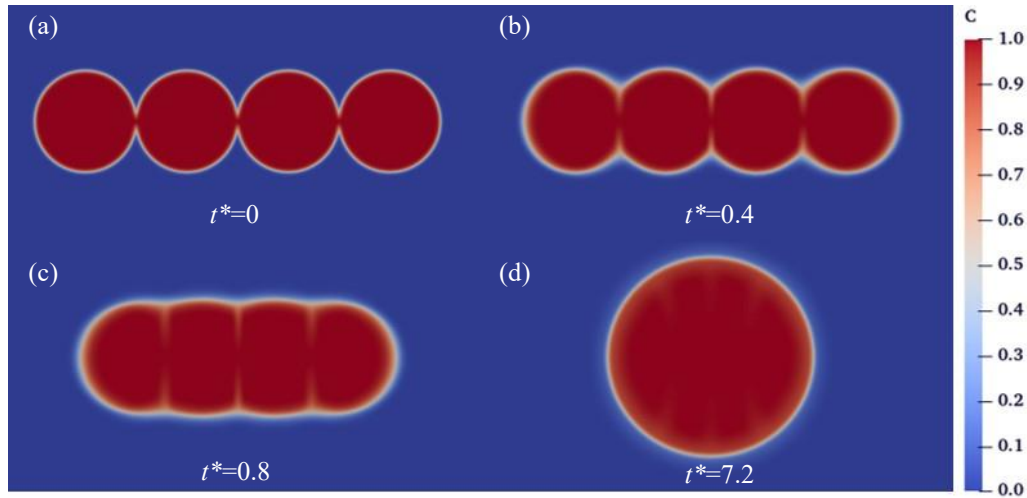

**Fig. 10.** The microstructural evolution of the chain model consisting of four equally-sized particles at different times: (a) $t^*=0$, (b) $t^*=0.4$, (c) $t^*=0.8$, and (d) $t^*=7.2$.

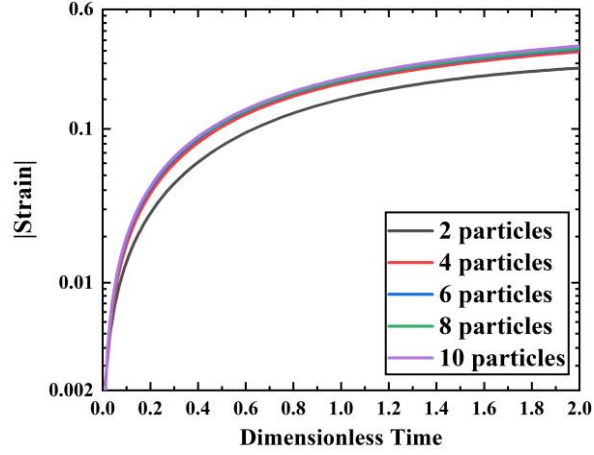

**Fig. 11.** The strain evolutions of the chain models consisting of different numbers of equally sized particles.

To further investigate the impact of particle size and arrangement on viscous sintering, the 4-particle chain model with two different particle sizes and four different arrangements are designed, as shown in Fig. 12. The $x$-strain evolution of each design is shown in Fig. 14a. Initially, the strains for the arrangements in Fig. 12 show no significant difference for $t^*\leqslant 0.6$. In the subsequent sintering process, the design b in Fig. 12 exhibits the fastest sintering speed, followed by designs a and c, while the design d shows the slowest sintering speed. This is due to that although all the four designs in Fig. 12 have the same surface energy initially, the design d is closer to the finial equilibrium state, i.e., a large circle with conserved area. As such, the design d will have faster energy dissipation, as shown in Fig. 14b. Eventually, all four curves gradually converge as the four designs share the same equilibrium state.

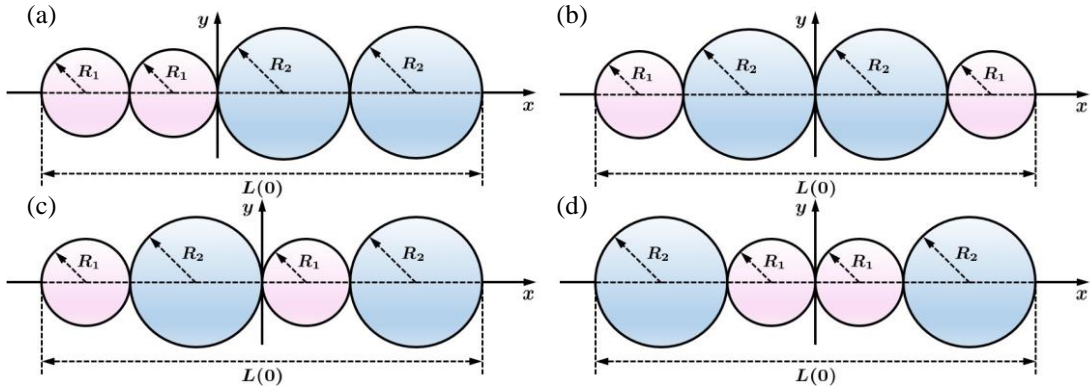

**Fig. 12.** The four-particle chain model consisting of two different particle sizes and four different arrangements, with a size ratio of $R_2/R_1=2$.

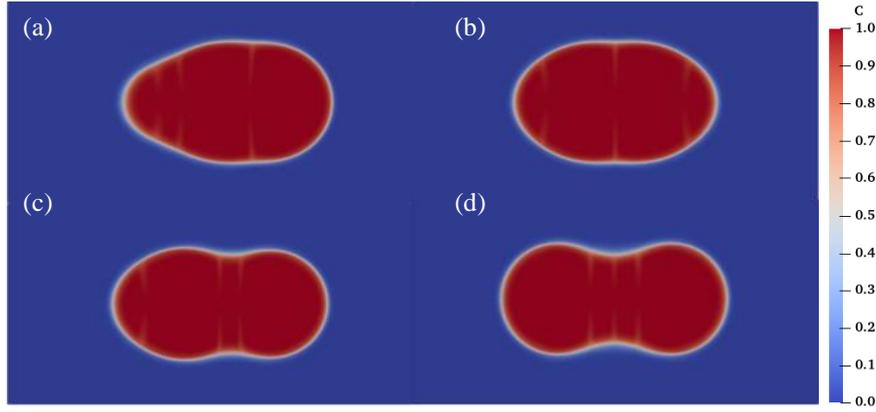

**Fig. 13.** The morphologies of the four-particle models of different arrangements when the normalized time $t^*= 3$.

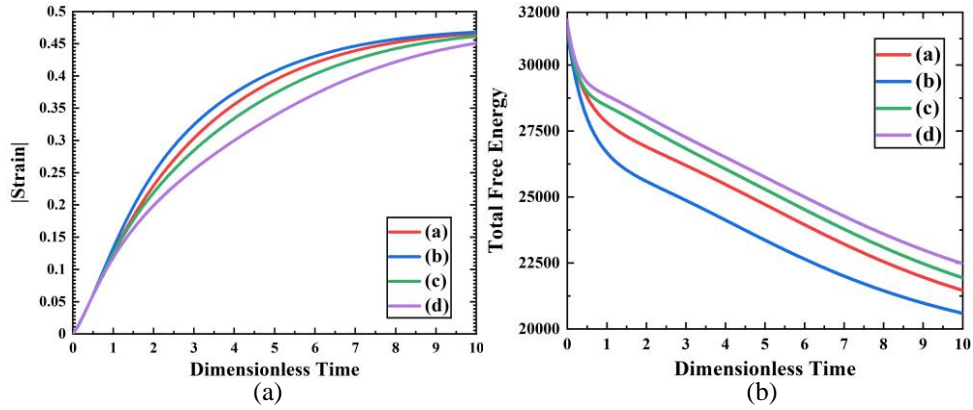

**Fig. 14.** The $x$-strain (a) and total free energy (b) evolutions as a function of normalized time for the four-particle models with different arrangements.

4.3. Particle shape effect

To investigate the influence of particle shapes on the sintering process, we implemented viscous sintering models with various elliptical particle shapes, as illustrated in Fig. 15. In these models, the major axis is denoted by $A$ and the minor axis by $B$, with the axis ratio defined as $\lambda=B/A\leqslant 1$. To ensure comparability across models, the total area of each ellipse was kept constant, thereby maintaining the same equilibrium state for all configurations. Notably, $\lambda = 1$ corresponds to the sintering of two equal-sized circles. Fig. 16 depicts the morphology of each elliptical model at the normalized time $t^*=3$, where time normalization is based on the radius in the case of $\lambda = 1$.

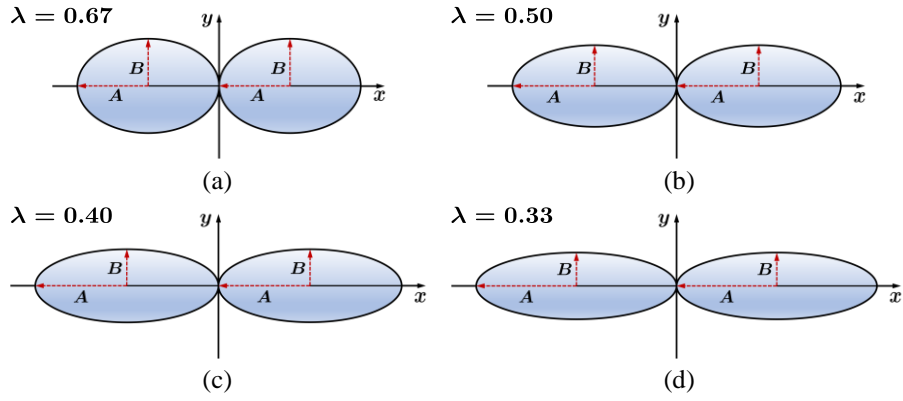

**Fig. 15.** The initial shapes of the ellipse models for different values of $\lambda$.

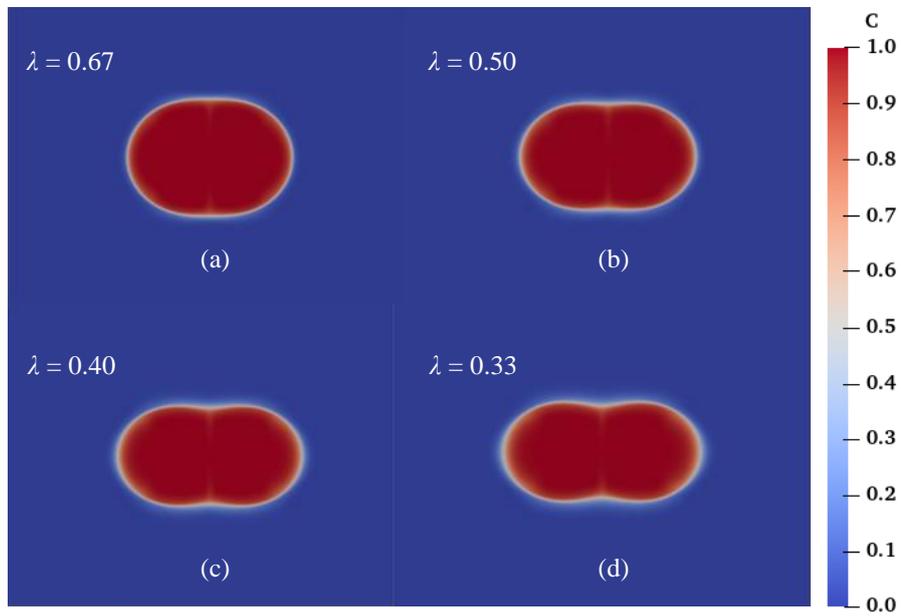

**Fig. 16.** The morphologies for the four models with different $\lambda$ values when the normalized time $t^* = 3$.

The $x$-Strain and total free energy evolutions of different ellipse models with different $\lambda$ values are shown in Fig. 17. On the one hand, the strain rate decreases when $\lambda$ goes up. This is due to that the elliptic models with larger $\lambda$ have smaller free energy initially (as shown in Fig. 17b). As all the elliptic models share the same equilibrium by design, models with smaller free energy initially will generate less driving force for the shrinkage. However, the time required to reach the equilibrium strain is also decreasing when $\lambda$ rises. As models with larger $\lambda$ always have smaller energies, their morphologies (as shown in Fig. 16) are closer to the equilibrium and thus requires less time to finish sintering.

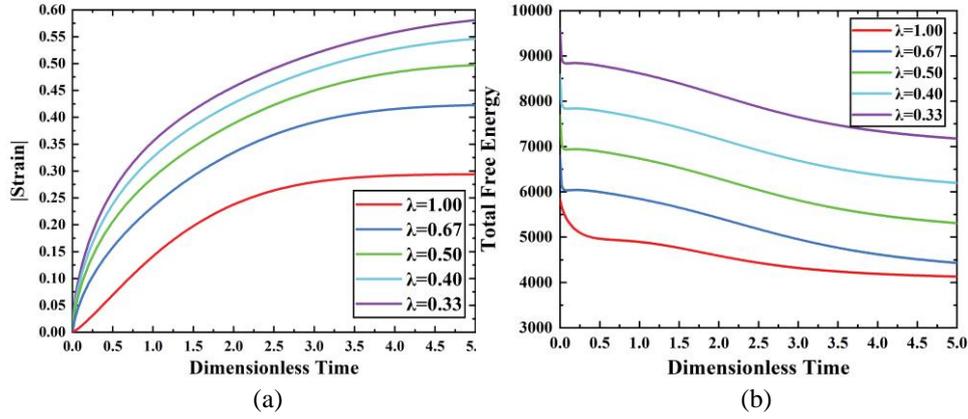

**Fig. 17.** The *x*-strain (a) and total free energy (b) evolutions as a function of normalized time for the models with different $\lambda$ values.

4.4. Stress distribution

This study examines the stress distribution during the viscous sintering process of linear chains composed of particles of the same size and those with varying sizes. The analysis focuses on the evolution of stress and its role in driving morphological changes. Fig. 18 plots the Frobenius norm of the stress for each model at different time steps, highlighting the significant stress concentrations around the neck regions, attributed to the local curvature. These localized stresses create gradients in the contact area, facilitating an increase in the contact radius between particles through viscous flow mechanisms. Concurrently, the stress gradient along the *x*-axis drives a reduction in the chain's overall length, indicating a progressive densification of the structure.

As sintering advances, these stress gradients diminish, leading to a more uniform stress distribution across the chain. This transition is particularly evident in the later stages (e.g., at $t^* = 12.0$), where the stress patterns stabilize, reflecting a homogenized internal structure. Interestingly, the observed trends hold consistently for all examined models, irrespective of particle size uniformity, suggesting that the underlying mechanics are robust across different geometries. These findings underline the critical role of stress gradients in shaping the microstructural evolution during sintering and may offer insights into optimizing material properties through tailored particle arrangements and stress management.

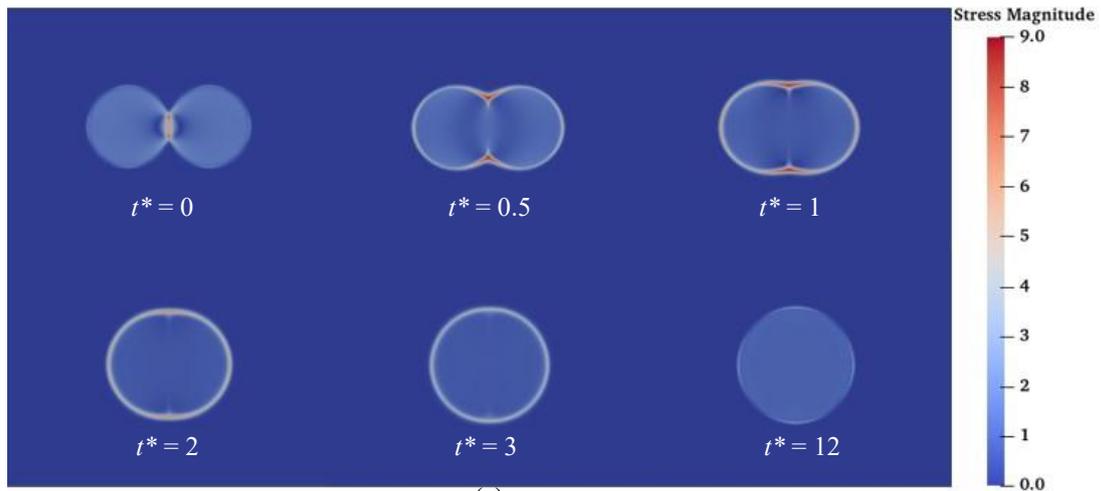

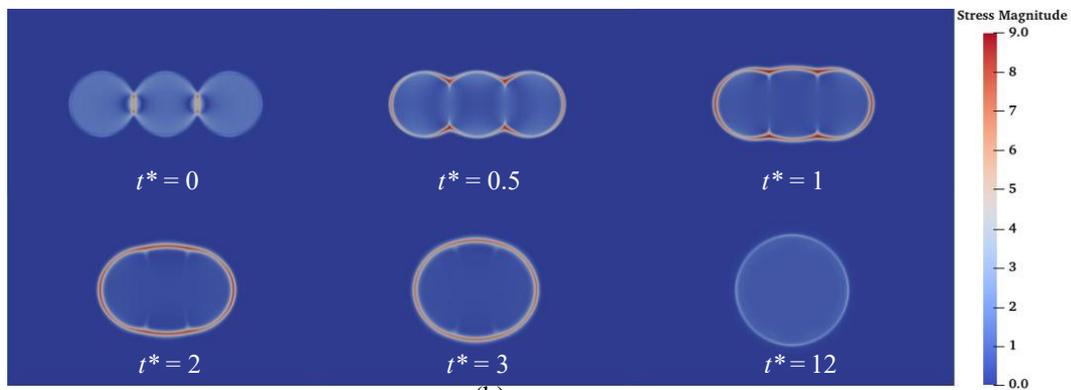

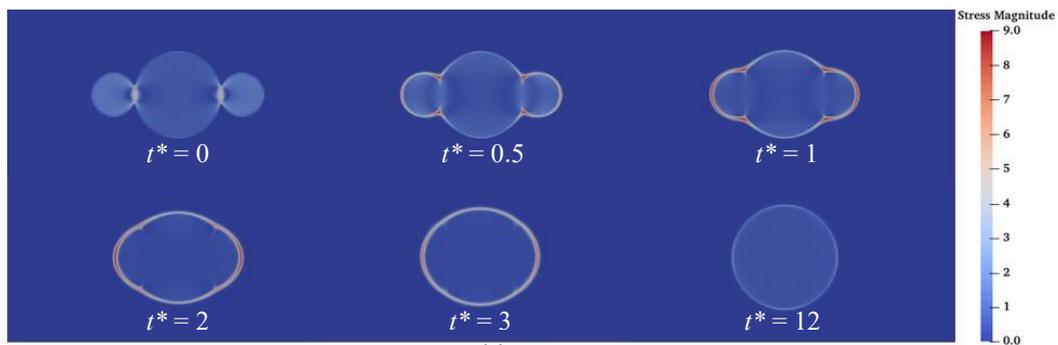

**Fig. 18.** The distribution of Frobenius norm of stress within different particle systems at different times: (a) the two-particle model; (b) the chain model consisting of three identical particles; (c) the chain model consisting of three particles with varying sizes; $t^*$ represents the dimensionless time normalized with respect to the radius of the smaller particle.

## 5. Summary and conclusions

This study introduces a thermodynamically consistent phase-field-micromechanics model for viscous sintering considering dramatic viscosity change across the particle-medium interface. The model formulates the governing equations by incorporating the energetic variational principle and allows for the implicit description of particle morphology evolution without necessitating the development of specialized numerical techniques for explicit interface tracking. As such, the introduced model facilitates numerical implementation. The validity of the proposed model is examined through qualitative and quantitative comparisons with analytical solutions and experimental measurements for two materials, PA12 and ABS, under both isothermal and non-isothermal conditions.

Building on this foundation, the model was applied to a range of numerical simulations to investigate the effects of particle size, shape, and arrangement on the sintering process. Key phenomena such as contact radius growth, strain evolution, and stress distribution during viscous sintering were analyzed in detail. Looking forward, the phase-field-micromechanics model will be extended to three-dimensional and multi-particle systems to explore the sintering characteristics of representative volume elements for real material systems. Furthermore, the model's inherent universality and scalability pave the way for broader applications, including multi-physics and multiphase sintering processes. Additionally, the integration of machine learning and artificial intelligence technologies holds promise for data-driven advancements, enabling seamless fusion of experimental data and sintering simulations for predictive modeling and optimization.

**Declaration of Competing Interest**

The authors declare that they have no known competing financial interests or personal relationships that could have appeared to influence the work reported in this paper.

**Acknowledgements**

Q.Y. acknowledges the financial support from the Science and Technology Commission of Shanghai Municipality through grant No. 23010500400, the key program (No.

52232002), and the grant (No. 12272214) from the Natural Science Foundation of China.